\documentclass[letterpaper, 10 pt, conference]{ieeeconf}  

\IEEEoverridecommandlockouts                                  
\usepackage{tabularray}
\usepackage{cite}
\usepackage{amsmath,amssymb,amsfonts}
\usepackage[]{footmisc}
\usepackage{xcolor}
\usepackage{algorithm}
\usepackage{algorithmic}
\usepackage{graphicx}
\usepackage{textcomp}
\usepackage{psfrag}
\usepackage{epsfig}
\usepackage{subcaption}
\usepackage{float}
\usepackage{xcolor}
\usepackage{bm}
\usepackage{tabularx}
\usepackage{booktabs}
\newtheorem{mydef}{ Definition}
\newtheorem{mypro}{Proposition}

\newtheorem{myrem}{Remark}

\newtheorem{mythr}{Theorem}

\newtheorem{myexa}{Example}

\begin{document}

\title{\LARGE Direct Shooting Method for Numerical Optimal Control: A Modified Transcription Approach}

\author{Jiawei Tang, Yuxing Zhong, Pengyu Wang, Xingzhou Chen, Shuang Wu,~and~Ling~Shi,~\IEEEmembership{Fellow,~IEEE}
	\thanks{Jiawei Tang, Yuxing Zhong, Pengyu Wang, Xingzhou Chen, and Ling Shi are with the Department of Electronic
		and Computer Engineering, Hong Kong University of Science and
		Technology, Clear Water Bay, Hong Kong SAR (email: jtangas@connect.ust.hk; yzhongbc@connect.ust.hk; pwangat@connect.ust.hk; xchenfk@connect.ust.hk; eesling@ust.hk).}
  \thanks{Pengyu Wang is also with Shenzhen Key Laboratory of Robotics Perception and Intelligence and the Department of Electronic and Electrical Engineering, Southern University of Science and Technology, Shenzhen, China.}
   \thanks{Shuang Wu is with the Noah's Ark Lab, Huawei (email:  wushuang.noah@huawei.com).}
	}
\newif\ifarxiv

\arxivfalse
\arxivtrue

\maketitle

\begin{abstract}
Direct shooting is an efficient method to solve numerical optimal control. It utilizes the Runge-Kutta scheme to discretize a continuous-time optimal control problem making the problem solvable by nonlinear programming solvers. However, conventional direct shooting raises a contradictory dynamics issue when using an augmented state to handle {high-order} systems. This paper fills the research gap by considering the direct shooting method for {high-order} systems. We derive the modified Euler and Runge-Kutta-4 methods to transcribe the system dynamics constraint directly. Additionally, we provide the global error upper bounds of our proposed methods. A set of benchmark optimal control problems shows that our methods provide more accurate solutions than existing approaches.
\end{abstract}
\IEEEpeerreviewmaketitle

\section{Introduction}
Direct transcription methods play a crucial role in numerical approaches for solving optimal control problems. They convert the continuous-time problem into a finite-dimensional one through discretization so that the optimal control trajectory can be computed using nonlinear programming (NLP) solvers. Because of the flexibility to handle various types of systems and constraints, direct transcription methods can be adapted to different real-world applications \cite{yang2022hierarchical,mpcc-uzh,10124823}. Moreover, the abundance of useful monographs\cite{betts2010practical,kelly2017introduction} and open-source software \cite{Kelly_OptimTraj_Trajectory_Optimization_2022, GPOPS-II,becerra2010psopt} facilitate the widespread use of direct transcription methods.

Direct transcription methods can be categorized into direct collocation and direct shooting. Unlike direct collocation, which parameterizes the control trajectory and the state trajectory simultaneously using a set of collocation points, direct shooting only uses control parameterization. The states are implied by integrating the dynamics forward in time. Besides, direct shooting offers an advantage compared to direct collocation. It enables us to delve into the Markov structure of discrete-time optimal control problems and facilitates the development of fast optimization techniques, such as differential dynamic programming (DDP) \cite{mayne1973differential} and iterative linear quadratic regulator (ILQR) \cite{li2004iterative}. The well-developed fast numerical optimal control solvers \cite{altro2019, OCS2, PWA} relying on this advantage make direct shooting rapidly popular in diverse applications, such as autonomous driving \cite{junma_admm}, mobile vehicles \cite{Alcan2023TrajectoryOO}, and quadrupedal robots \cite{9811647}.


Direct shooting will reduce accuracy if the approximation schemes used in problem transcription are not chosen appropriately. A typical issue is the contradictory dynamics when dealing with {high-order} system dynamics. This issue exists in all direct transcription methods but was mostly ignored until the recent research on direct collocation for second-order systems \cite{moreno2022collocation, simpson2022direct,moreno2022legendre}. They found that using the augmented state to transform the second-order ordinary differential equation to a first-order one will introduce additional numerical error. To solve this issue, the second-order trapezoidal and second-order Hermite-Simpson methods were introduced in \cite{moreno2022collocation}. Further, Simpson et al. \cite{simpson2022direct} and Martin et al. \cite{moreno2022legendre} extended the idea to the global collocation method and the Legendre-Gauss pseudospectral collocation method, respectively.

Different from existing works, our research focuses on direct shooting. The direct collocation relies on the function approximation for problem transcription, which cannot be applied to the shooting method directly. Besides, the above works only demonstrated the effectiveness of their methods in numerical examples. These factors motivate our research.


This paper investigates the direct shooting method for
{high-order} systems. Here are our contributions:

\textcolor{black}{1) We evaluate the contradictory dynamics issue of the direct shooting for {high-order} systems and propose the modified Euler and Runge-Kutta-4 (RK4) methods to address the issue.}

2) We provide the global error upper bounds the proposed modified shooting methods (\textit{Theorem 1} and \textit{Theorem 2}), addressing the lack of convergence analysis in recent numerical schemes for {high-order} systems.

\textcolor{black}{3) We evaluate our proposed methods with several benchmark optimal control problems, and the numerical results illustrate the superior performance of the proposed methods.}

{\it Notations}:  The notation $\mathbb{R}^{n}$ denotes the set of real vectors with  $n$ elements. The $i$-th element of a vector $ {v}\in \mathbb{R}^{n}$ is denoted by $[ {v}]_i$. The notation $t_k$ denotes the time at knot point $k$. The notation $ {x}_k =  {x}(t_k)$ and $ {u}_k =  {u}(t_k)$ denote the state and control at knot point $k$, respectively. We use $ {\dot{q}}(t)=\frac{d}{dt} {q}(t)$, $ {\ddot{q}}(t)=
\frac{d^2}{dt^2} {q}(t)$, and $q^{(i)}(t) = \frac{d^i}{dt^i}q(t)$ to denote the first-order, second-order and the $i$th-order time derivative of ${q}(t)$. We use interval notation $j \in [a, b):=\{a, a+1, \dots, b-1\}$, $j \in [a, b]:=\{a, a+1, \dots, b\}$, for $a, b\in \mathbb{N}$ to denote the sets of consecutive integers. The norm $\|\cdot\|$in this paper is assumed to be Euclidean if not specified.
\section{Numerical Optimal Control}
\subsection{Nonlinear Optimal Control}
Consider a general nonlinear system
\begin{equation} 
    {\dot{x}}(t) = {f}_1({x}(t), {u}(t)),\label{1stDyn1}
\end{equation}
where ${x}(t) \in \mathbb{R}^{n_x} $ is the state and ${u}(t)  \in \mathbb{R}^{n_u}$ is the control input. The performance index for (\ref{1stDyn1}) is defined as
\begin{equation}\label{ctObj}
    J = \phi({x}(0), {x}(t_f)) + \int_{0}^{t_f}l({x}(\tau), {u}(\tau))d\tau,
\end{equation}
{where $\phi(\cdot)$  is the terminal cost function and $l(\cdot)$ is the intermediate cost function. Given a general inequality constraint $ {g}( {x}(t),  {u}(t)) \leq  {0}$ and a boundary equality constraint $ {b}( {x}(0),  {x}(t_f), t_f) =  {0}$, the problem to be solved is presented as follows.}

\noindent\textit{\textbf{Problem 1}: (General Optimal Control Problem)}
\begin{subequations}
\begin{align}
     \min_{ x(\cdot), {u}(\cdot)} &\phi(  {x}(0),  {x}(t_f)) +\int_{0}^{t_f}l( {x}(\tau),  {u}(\tau))d\tau \label{p1_a}\\
    \text{s.t.} ~~~ & {g}( {x}(t),  {u}(t)) \leq  {0},~~0 \leq t \leq t_{f}, \label{p1_b}\\     
    & {b}( {x}(0),  {x}(t_f), t_f) =  {0},\label{p1_c}\\
    &  {\dot{x}}(t) = {f}_1({x}(t), {u}(t)), ~~0 \leq t \leq t_{f}. \label{p1_d}
\end{align}
\end{subequations}
\subsection{First-order Direct Shooting}
\textcolor{black}{\textit{\textbf{Problem 1}} is difficult to solve because it involves infinite-dimensional optimization. Classical methods require deriving optimal conditions based on the calculus of variations and solving them indirectly\cite{betts2010practical}.} The direct shooting method utilizes the discretization technique to convert \textit{\textbf{Problem 1}} into a finite-dimensional optimization problem. In particular, the continuous state and control functions are approximated by discrete sets of real numbers, known as knot points. In particular, for $0 \leq t \leq t_f$, we have 
\begin{align*}
t &\rightarrow t_0, \ldots, t_k, \ldots,t_N,\\
{x}(t) &\rightarrow {x}_0 , \ldots, {x}_k, \ldots,{x}_N,\\
{u}(t) &\rightarrow {u}_0 , \ldots, {u}_k, \ldots, {u}_{N},
\end{align*}  
where $x_k$ and $u_k$ are the approximations to $x(t_k)$ and $u(t_k)$, respectively. With the initial condition $x(0) = x_0$, direct shooting builds state propagation equation based on Runge-Kutta scheme\cite{betts2010practical}, i.e., $\forall k\in [0, N)$,
    \begin{subequations}\label{1st-RK-SCHEME}
    \begin{align}
    &{x}_{k+1} = {x}_{k} + \sum_{i=1}^{s}b_i K_i,\\
    &K_i = h \cdot f_1(x_k+\sum_{j=1}^{i-1}a_{i,j}K_j, u_k),
\end{align}
\end{subequations}
where $h = t_{k+1} - t_k$ is the step size, ${a}_{i,j}\in \mathbb{R}$, ${b}_i \in \mathbb{R}$ are coefficients determined by Taylor theorem, and $s$ is the stage of Taylor expansion. The well-known Euler method and the Runge-Kutta-4 (RK4) method are with $s=1$ and $s=4$, respectively. 

By utilizing the Runge-Kutta scheme, we can convert the decision variables from functions to real numbers. Moreover, we can convert the differential equation constraints (\ref{p1_d}) into equality constraints and convert the integral in the objective function (\ref{p1_a}) into a summation accordingly. Besides, through enforcing the inequality constraint (\ref{p1_b}) and the boundary equality constraint (\ref{p1_c}) at each knot point, \textit{\textbf{Problem 1}} is converted into a finite-dimensional optimization problem, which can be solved using off-the-shelf NLP solvers.
\subsection{Downside of First-order Direct Shooting Method}
{It should be noted that \textit{\textbf{Problem 1}} considers the first-order nonlinear system. However, many practical control systems are in a high-order form, i.e.,
\begin{equation} \label{highDyn}
q^{(\mathcal{N})}(t) = {f}_{\mathcal{N}}({q}(t), q^{(1)}(t),\cdots, q^{(\mathcal{N}-1)}(t), {u}(t)),
\end{equation}
where ${q}(t) \in \mathbb{R}^{n_q} $ is the system configuration and $\mathcal{N}$ determines the order. In order to solve the optimal control problem for the high-order system (replacing (\ref{p1_d}) with (\ref{highDyn}) in \textit{\textbf{Problem 1}}) using the first-order direct shooting method, some proposed to use following transformation:}

{\textit{Transformation 1:} The system dynamics (\ref{highDyn}) is cast into a first-order form using the augmented state ${x}(t) = ({q}(t), q^{(1)}(t),\cdots, q^{(\mathcal{N}-1)}(t))$, i.e.,
    \begin{equation*}\label{1stDyn}
        {\dot{x}}(t)= \begin{bmatrix}
        q^{(1)}(t)\\
        q^{(2)}(t)\\
        \vdots\\
        q^{(\mathcal{N})}(t)
        \end{bmatrix}={f}_1({x}(t), {u}(t)) =\begin{bmatrix}
        q^{(1)}(t)\\
         q^{(2)}(t)\\
        \vdots\\
        {f}_{\mathcal{N}}({q}(t), {\cdots}, {u}(t))
        \end{bmatrix}.
    \end{equation*}}

\textit{Transformation 1} is widely used in control and robotics \cite{altro2019, OCS2, PWA}. However, combining it with (\ref{1st-RK-SCHEME}) leads to a contradiction. We use the following example to illustrate it.

\begin{myexa}
    Consider a linear second-order system
    \begin{equation} \label{lnear-2nd-sys}
        \ddot{q}(t) = u(t).
    \end{equation}
    Since the system is linear, we can use the Euler method ($s = 1$ in (\ref{1st-RK-SCHEME})) to handle the differential equation constraint. Following with the Euler method with \textit{Transformation 1}, the differential equation constraint (\ref{lnear-2nd-sys}) is converted into the following equality constraints:
    \begin{subequations}\label{linear_u_1st_all}
        \begin{align}
            {q}_{k+1} &= {q}_{k} +\dot{q}_kh , ~~ k \in [0, N)\label{linear_u_1st_1},\\
        \dot{q}_{k+1} &= \dot{q}_{k} +  u_kh ~~ k \in [0,N). \label{linear_u_1st_2}
        \end{align}
    \end{subequations}
    However, the analytical state propagation equation is
    \begin{subequations}\label{analytical_linear_trans}
        \begin{align}
            {q}_{k+1} &= {q}_{k} + \dot{q}_k h + \frac{1}{2} u_k h^2, ~~ k \in [0, N),\\
        \dot{q}_{k+1} &= \dot{q}_{k} +  u_kh ~~ k \in [0,N) .
        \end{align}
    \end{subequations}
    
    It is easy to see that some transcription error is introduced to (\ref{linear_u_1st_all}). (\ref{linear_u_1st_all}) approximates both $q(t)$ and $\dot{q}(t)$ as linear function between adjacent knot points. However, considering the inherent mathematical relationship between $q(t)$ and $\dot{q}(t)$, $q(t)$ should be quadratic if $\dot{q}(t)$ is linear.
\end{myexa}

The above discussion indicates that the combination of  \textit{Transformation 1} with (\ref{1st-RK-SCHEME}) reduces the approximation accuracy. Therefore, it is critical to consider the inherent relationship of $q(t)$ and its time derivatives when designing the numerical scheme for the high-order system. In the next section, modified direct shooting methods are proposed to alleviate the aforementioned issues of the conventional first-order direct shooting method.

\section{Modified Direct Shooting}
{In this section, we present two modified direct shooting methods: the Euler method and the RK4 method. Instead of utilizing \textit{Transformation 1}, we derive the state propagation equation from the system dynamics equation. To provide a clear illustration, we focus on the second-order system, i.e.,
\begin{equation} \label{2ndDyn}
{\ddot{q}}(t) = {f}_2({q}(t), {\dot{q}}(t), {u}(t)).
\end{equation}
It serves as a key example to explain the fundamental concept behind our proposed method. Subsequently, we will discuss the extension of this approach to the high-order system \eqref{highDyn}.}
\subsection{Second-order Euler Method}
\begin{mypro}
    Under the first-stage Runge-Kutta scheme, the second-order differential equation constraint (\ref{2ndDyn}) is equivalent to the following equality constraints, i.e., $\forall k \in [0, N)$,
    \begin{subequations}\label{2nd-Euler}
        \begin{align}
        {q}_{k+1} &= q_k + h\dot{q}_k + \frac{1}{2} h K_1, \label{2nd-euler-q}\\
        \dot{q}_{k+1}& = \dot{q}_k + K_1,\label{2nd-euler-qdot}\\
        K_1 &= h \cdot f_{2}(q_k,\dot{q}_k, u_k).
        \end{align}
    \end{subequations}
    \noindent \textit{Proof:} For $t_k \leq t \leq t_{k+1}$, the Euler method assumes that $\dot{q}(t)$ is approximated by the first-order Taylor polynomial around knot point $k$. Hence, we have
\begin{equation*}
    \dot{q}(t) = \dot{q}_k +  \ddot{q}_k(t-t_k). \label{eq_v}
\end{equation*}
Through  writing  $q(t)$ in the integral form, we have
\begin{align*}
q(t) &= q_k+\int_{t_k}^{t} \dot{q}(\tau)~d\tau,\\
    {q}(t) &= q_k + \dot{q}_k (t-t_k) + \frac{1}{2} \ddot{q}_k (t-t_k)^2.
\end{align*}
As $\ddot{q}_k = f_2(q_k,\dot{q}_k,u_k)$, (\ref{2nd-Euler}) is directly followed from it. The transcription of (\ref{2ndDyn}) by Euler method is completed.\hfill $\square$
\end{mypro}

Note that (\ref{2nd-Euler}) builds state propagation equations for the second-order system. In this case, the first-order Taylor series approximation only applies to the first-order derivative of the configuration \eqref{2nd-euler-qdot}, while the configuration propagation \eqref{2nd-euler-q} is calculated based on the integral relationship between $q(t)$ and $\dot{q}(t)$. It is worth mentioning that with the formulation shown in (\ref{2nd-Euler}), the control input $u_k$ takes effect on the next configuration $q_{k+1}$, which solves the delay issue of the first-order method as explained in the \textit{Example 1}.

\subsection{Second-order RK4 Method}
The Euler method is a simple and straightforward numerical method to handle the differential equation. However, it has a larger truncation error compared to other numerical methods\cite{burden2015numerical}, which means that the accuracy of the solution decreases rapidly as the step size $h$ increases. In the following section, we will introduce the RK4 method for the second-order system. It is a more accurate and widely used numerical scheme in real-world robotic applications.
\begin{mypro}
    Under the fourth-stage Runge-Kutta scheme, the second-order differential equation constraint (\ref{2ndDyn}) is equivalent to the following equality constraints, i.e., $\forall k \in [0,N)$,
   \textcolor{black}{
     \begin{subequations} \label{2nd-RK4}
   \begin{align}
   {q}_{k+1} &= q_k + h\dot{q}_k + \frac{h}{5}K_1 + \frac{h}{6}K_2+\frac{h}{10}K_3+\frac{h}{30}K_4,\\
    \dot{q}_{k+1}& = \dot{q}_k + \frac{1}{6}K_1 + \frac{1}{3}K_2+\frac{1}{3}K_3+\frac{1}{6}K_4,\\
    K_1 &= h \cdot f_{2}( q_k, \dot{q}_k, u_k).\label{k_1}\\
    K_2 &= h \cdot f_2( q_k + \frac{h}{2}\dot{q}_k, \dot{q}_k + \frac{K_1}{2},u_k),\\
    K_3 &= h \cdot f_{2}(q_k +\frac{h}{2}\dot{q}_k, \dot{q}_k + \frac{K_2}{2},u_k),\\
    K_4 &= h \cdot f_2(q_k+\dot{q}_k, \dot{q}_k + K_3,u_k).\label{k_4}
\end{align}
   \end{subequations}
   }

   
   \noindent \textit{Proof:} For $t_k \leq t \leq t_{k+1}$, the RK4 method assumes that $\dot{q}(t)$ follows the fourth-order Taylor polynomial around knot point $k$. Hence we have
\begin{equation}
    \begin{aligned}
       \dot{q}(t) &= \dot{q}_k + \sum_{i=1}^4\frac{1}{i!}(t-t_k)^if_{2,k}^{(i-1)}.
    \end{aligned}
\end{equation}
 The notation $f^{(i)}(\cdot)$ denotes the $i$th-order time-derivative of function $f(\cdot)$. Through writing $q(t)$ in the integral form, we have the following relationship:
\begin{equation*}
    {q}(t) = q_k + \dot{q}_k (t-t_k) +\sum_{i=1}^4\frac{1}{(i+1)!}(t-t_k)^{i+1}f_{2,k}^{(i-1)}.
\end{equation*}
Denote that 
    \begin{align*}
    K_1 &= h  f_{2,k}, ~~~K_2= h  f_{2,k} + \frac{h^2}{2}f_{2,k}^{(1)},\\
    K_3 &= h  f_{2,k} +  \frac{h^2}{2}f_{2,k}^{(1)} + \frac{h^3}{4} f_{2,k}^{(2)},\\
    K_4 &= h  f_{2,k} + {h^2}f_{2,k}^{(1)} + \frac{h^3}{2} f_{2,k}^{(2)}+ \frac{h^4}{4} f_{2,k}^{(3)}.
\end{align*}
We have
\begin{equation*}
    \begin{aligned}
        &\dot{q}(t_{k+1}) = \dot{q}_k + \frac{1}{6}hf_{2,k} +\frac{1}{3}(h f_{2,k}+\frac{h^2}{2} f_{2,k}^{(1)})\\
    &~~~~~~~~~~~~~~~~~~+\frac{1}{3}(h  f_{2,k} +  \frac{h^2}{2}f_{2,k}^{(1)} + \frac{h^3}{4} f_{2,k}^{(2)})\\
    &~~~~~~~~~~~~~~~~~~+\frac{1}{6}(h  f_{2,k} + {h^2}f_{2,k}^{(1)} + \frac{h^3}{2} f_{2,k}^{(2)}+ \frac{h^4}{4} f_{2,k}^{(3)})\\
            &~~~~~~~~~~~~= \dot{q}_k + \frac{1}{6}K_1 + \frac{1}{3}K_2+\frac{1}{3}K_3+\frac{1}{6}K_4,
    \end{aligned}
\end{equation*}
\begin{equation*}
    \begin{aligned}
        &{q}(t_{k+1}) = q_k + h \dot{q}_k +\frac{h^2}{5} f_{2,k} + \frac{h}{6}(h f_{2,k}+\frac{h^2}{2} f_{2,k}^{(1)}) \\
     &~~~~~~~~~~~~~~~~~~+\frac{h}{10}(h  f_{2,k} +  \frac{h^2}{2}f_{2,k}^{(1)} + \frac{h^3}{4} f_{2,k}^{(2)})\\
    &~~~~~~~~~~~~~~~~~~+\frac{h}{30}(h  f_{2,k} + {h^2}f_{2,k}^{(1)} + \frac{h^3}{2} f_{2,k}^{(2)}+ \frac{h^4}{4} f_{2,k}^{(3)})\\
    &~~~~~~~~~~~~= q_k + h \dot{q}_k + \frac{h}{5}K_1 + \frac{h}{6}K_2+\frac{h}{10}K_3+\frac{h}{30}K_4.
    \end{aligned}
\end{equation*}
By Taylor's theorem in multiple variables \cite{burden2015numerical}, we can obtain a compact form for $K_i$, as shown in (\ref{k_1})-(\ref{k_4}). Hence the transcription of (\ref{2ndDyn}) by RK4 method for the second-order system is completed.\hfill $\square$
\end{mypro}

Note that (\ref{2nd-RK4}) builds state propagation equations for the second-order system with a high-order Taylor series approximation. Similar to \textit{Statement 1}, the control input $u_k$ takes effect immediately to the next configuration $q_{k+1}$. Compared to the Euler method, the RK4 method utilizes a weighted average of the derivative estimations to achieve fourth-order Taylor series approximation on $\dot{q}(t)$, which results in higher numerical accuracy than the Euler method. The detailed convergence analysis of our proposed methods will be discussed in Section IV.
\subsection{Extension to the High-order System}
{The above results demonstrate the key idea of the proposed modification in the second-order system. To extend this idea to the general high-order system, one can use the relationship between successive orders of the time derivative of $q(t)$. For instance, within the time interval $t_k \leq t \leq t_{k+1}$, we have the following integral expressions.
\begin{align*}
q^{(\mathcal{N}-2)}(t) &= q^{(\mathcal{N}-2)}_k+\int_{t_k}^{t} q^{(\mathcal{N}-1)}(\tau)~d\tau,\\
q^{(\mathcal{N}-3)}(t) &= q^{(\mathcal{N}-3)}_k+\int_{t_k}^{t} q^{(\mathcal{N}-2)}(\tau)~d\tau,\\
& ~~\vdots \\
q(t) &= q_k+\int_{t_k}^{t} q^{(1)}(\tau)~d\tau.
\end{align*}
By employing the derivation of the Runge-Kutta scheme described in \textit{Proposition 1} and \textit{Proposition 2} and recognizing the above integral relationship, we can effectively extend the proposed idea to the high-order system.}
\section{Convergence Analysis}
In this section, we conduct the convergence analysis on the modified Euler and RK4 methods. To conduct analysis, we first define the numerical approximation error and the convergence condition.


\begin{mydef}
\textit{(Global Truncation Error\cite{burden2015numerical})} The global truncation error of the configuration approximation at time $t = t_k$ is defined as
\begin{equation*}
{e}_{k}(h) = {q}(t_{k})-q_k,
\end{equation*}
where ${q}(t_{k})$ is the exact solution of the configuration at $t = t_{k}$, $q_k$ is the approximation of the solution at $t = t_{k}$ with the condition $q_0={q}(0)$, and $h = t_{k+1} - t_{k}$.
\end{mydef}

\begin{mydef}
\textit{(Convergence Condition \cite{burden2015numerical})} The configuration approximation is said to be convergent with respect to the differential equation it approximates if
    \begin{equation*}
        \lim_{h \rightarrow 0} \max_{1\leq k \leq N}\|{e}_k(h)\| = 0.
    \end{equation*}    
\end{mydef}

By these definitions, we have the following results.

\begin{mythr} \textit{(Convergence of Second-order Euler Method)} The configuration approximation stated in (\ref{2nd-Euler}) is convergent with the global truncation error
\begin{equation}
    \|e_k(h)\| \leq \frac{\alpha h^3}{6hL+3h^2L^2}(e^{t_fL}-1), \label{tm3}
\end{equation}
if there exists a Lipschitz constant $L > 0$ with $\|q^{(i+1)}(t_1) - q^{(i+1)}(t_2)\|\leq L\|q^{(i)}(t_1) - q^{(i)}(t_2)\|$  for $0\leq t_1, t_2\leq t_f$ and $i \in [0,1]$, and a constant $\alpha > 0$ with $\|q^{(3)}(t)\|\leq \alpha$ for $0\leq t\leq t_{f}$.

\ifarxiv
\noindent \textit{Proof:} As ${e}_k(h) = {q}(t_{k})-q_k$, for $0 \leq \eta \leq t_f$, we have
\begin{equation*}
    \begin{aligned}
        \|{e}_k(h)\|&= \|{q}(t_{k-1})+h{\dot{q}}(t_{k-1})+\frac{h^2}{2}{\ddot{q}}(t_{k-1})+\frac{h^3}{6}{q}^{(3)}(\eta)\\
    &-({{q}}_{k-1} + h{{\dot{q}}}_{k-1}  + \frac{1}{2}h^2\ddot{q}_k)\|.\\
    \end{aligned}
\end{equation*}
Based on triangle inequality, we have
\begin{equation*}
    \begin{aligned}
   \|{e}_k(h)\| &\leq \|{q}(t_{k-1})-{q}_{k-1}\|+h\|\dot{q}(t_{k-1})-{\dot{q}}_{k-1}\|\\
    &+ \frac{h^2}{2}\|\ddot{q}(t_{k-1})-{\ddot{q}}_{k-1}\| + \frac{h^3}{6} \alpha \\
    &\leq (1+hL+\frac{h^2L^2}{2})\|{e}_{k-1}(h)\|+ \frac{h^3}{6} \alpha .
    \end{aligned}
\end{equation*}
By the discrete Gronwall's Lemma, we have
\begin{equation*}
\begin{aligned}
   \|{e}_k(h)\| &\leq (1+hL+\frac{h^2L^2}{2})^k \|e_0(h)\|\\
   &+\frac{h^3}{6} \alpha  \frac{(1+hL+\frac{h^2L^2}{2})^k-1}{hL+\frac{h^2L^2}{2}}.
   \end{aligned}
\end{equation*}
Since $\|e_0(h)\|=0$ and $1+hL+\frac{h^2L^2}{2}\leq e^{hL}$ for all $h\geq 0$, we conclude that
\begin{equation*}
   \|{e}_k(h)\| \leq  \frac{\alpha h^3 }{6hL+3h^2L^2}(e^{t_fL}-1).
\end{equation*}
As $h\rightarrow 0$, $\lim_{h \rightarrow 0} \max_{1\leq k \leq N}\|{e}_k(h)\| = {0}$. Hence, (\ref{2nd-Euler}) is convergent. The proof is completed. \hfill $\square$
\else
\noindent\textit{Proof:} See xxx \hfill $\square$
\fi

\end{mythr}

\begin{mythr}\textit{(Convergence of Second-order RK4 method)} The configuration approximation stated in (\ref{2nd-RK4}) is convergent with the global truncation error
\begin{equation}
    \|e_k(h)\| \leq \frac{\beta h^6}{720\sum_{i=1}^5\frac{1}{i!}h^iL^i}(e^{t_fL}-1), \label{tm3}
\end{equation}
if there exists a Lipschitz constant $L > 0$ with $\|q^{(i+1)}(t_1) - q^{(i+1)}(t_2)\|\leq L\|q^{(i)}(t_1) - q^{(i)}(t_2)\|$  for $0\leq t_1, t_2\leq t_f$ and $i \in [0,4]$, and a constant $\beta > 0$ with $\|q^{(6)}(t)\|\leq \beta$ for $0\leq t\leq t_{f}$.

\ifarxiv
\noindent\textit{Proof:} The proof is similar to that of \textit{Theorem 1} and thus is omitted here. \hfill $\square$
\else 
\noindent\textit{Proof:} See xxx \hfill $\square$
\fi

\end{mythr}  
{\begin{myrem}
    For notational clarity, we define the Lipschitz continuity with the same constant $L$ for all $q^{(i)}$. 
\end{myrem}}
\begin{myrem}
We assume $q(t) \in C^{3}[0, t_f]$ for the Euler method and $q(t) \in C^{6}[0, t_f]$ for the RK4 method. Though this assumption cannot be guaranteed for all control systems, widely used practical dynamics, such as the unicycle, bicycle, and quadrotor, satisfy the assumption.
\end{myrem}


The above results show that the proposed modified Euler and RK4 methods converge. We can use the theoretical results on global truncation error bound to estimate the accuracy of the differential equation approximation, which can further benefit the estimation of the accuracy of the numerical solution and the implementation of mesh refinement.

\section{Numerical Experiments}
To evaluate the performance of the proposed methods, we compare the proposed modified shooting methods with the conventional ones on a number of benchmark optimal control problems for second-order systems. We choose the following four methods\footnote{1st-Euler and 1st-RK4 are widely used in existing numerical optimal control frameworks, such as ALTRO\cite{altro2019}, OCS2\cite{ OCS2}, and PWA \cite{PWA}.}.
\begin{itemize}
\item[1)] 1st-Euler: the Euler method with \textit{Transformation 1}.
\item[2)] 2nd-Euler: the modified Euler method in \textit{Proposition~1}.
\item[3)] 1st-RK4: the RK4 method with \textit{Transformation 1}.
\item[4)] 2nd-RK4: the modified RK4 method  in \textit{Proposition~2}.
\end{itemize}

 The problems are implemented in MATLAB with the symbolic framework CasADi\cite{andersson2019casadi} and the NLP solver IPOPT\cite{wachter2006implementation}. Each problem minimizes a quadratic objective and is subject to initial and terminal state constraints. In each problem, the final time $t_f$ is a fixed value.
\subsection{Problem Descriptions}
\begin{figure}[t]
	\centering
	\includegraphics[width=8.cm]{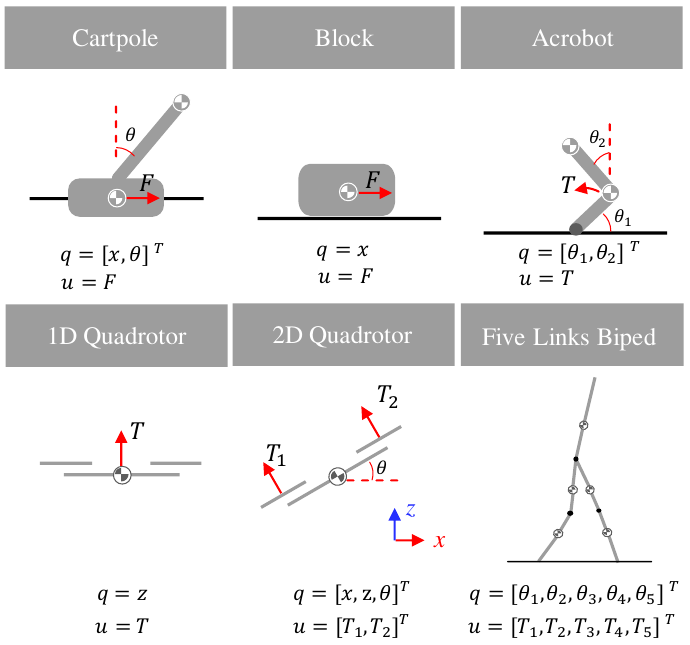}
	\caption{Schematics, configuration, and input vectors of the benchmark system dynamics.}
	\label{fig:figure1}
\end{figure}
Fig. \ref{fig:figure1} illustrates the schematics, configuration vector, and input vector of the benchmark system dynamics. The tasks to be solved are described as follows.
\begin{itemize}
\item[1)] \textit{Cartpole ($N=100$):} a pole attached to a cart via an unactuated joint. The cart can move along a frictionless track. The task is to swing the pole from its downward equilibrium position to its upward equilibrium position while adhering to certain control limits.
\item[2)] \textit{Block ($N=50$):} double integrator with one configuration. The task is to move the block with one meter.
\item[3)] \textit{Acrobot ($N=50$):} double pendulum system with one actuation. The task is to swing the Acrobot from its downward position to its upward position.
\item[4)] \textit{1D Quadrotor ($N=50$):} simplified quadrotor model with one configuration. The task is to move the quadrotor with one unit of length while overcoming gravity.
\item[5)] \textit{2D Quadrotor ($N=30$):} simplified quadrotor model with three configurations and two control inputs. The system is tasked to move from the start pose to the target pose subject to control limits.
\item[6)] \textit{Five Links Biped ($N=100$):} simplified bipedal model with five links connected by revolute joints. The joints are actuated by torque motors. The detailed dynamics of the robot can be found in \cite{kelly2017introduction}, with the parameters of the model matching those of the RABBIT \cite{1234651}. The task is to optimize the robot's gait subject to control limits.
\end{itemize}

\subsection{Performance Metrics}
\subsubsection{Accuracy}
To compare the accuracy of the four methods on the six problems mentioned above, we define the following error metric
\begin{align}
    \varepsilon(t) &=  \hat{q}(t)-q^{*}(t), \label{configError}
\end{align}
where $q^{*}(t)$ is the optimal configuration trajectory, while $\hat{q}(t)$ is the configuration recovered from the solver result $\{q_k, \dot{q}_k, \ddot{q}_k\}_{k=0}^N$ using cubic splines. We set large $N$ for solvers to get the results and treat them as the optimal solutions for the error metric.
To evaluate the total transcription error in each time interval, we use the following expression to determine the accumulated error in each time interval:
\begin{equation*}
    \eta_k = \int_{t_k}^{t_{k+1}}|\sum_{i=1}^{n}([\varepsilon(\tau)]_i)|d\tau, 
\end{equation*}
where the integral can be computed using the Rhomberg quadrature\cite{burden2015numerical}. The total transcription error of a trajectory is noted as $\eta_{total} = \sum_{k=0}^{N-1}\eta_k$.
\begin{table}[t]
\begin{center}
\caption{Total Transcription Error Comparison}
\begin{tabular}{|l||l|l|l|l|}
\hline
{Problem $\backslash$ Method} & {1st-Euler} & {2nd-Euler} & {1st-RK4}& {2nd-RK4} \\ \hline
\textit{Cartpole}        &     2.73        & 2.02 & 1.35 & \textbf{1.29} \\ \hline
\textit{Block  }      &   0.032        & 0.024 & \textbf{0.024}& \textbf{0.024} \\ \hline
\textit{Acrobot}       &     0.682        & 0.291 & 0.283 & \textbf{0.274} \\ \hline
\textit{1D Quadrotor}       &     0.007        & \textbf{0.004} & \textbf{0.004} & \textbf{0.004} \\ \hline
\textit{2D Quadrotor}       &     6.612        & 3.345 & 2.780 & \textbf{2.778} \\ \hline
\textit{Five Links Biped}       &    0.0162        & 0.0169 & \textbf{0.006} & \textbf{0.006}\\ \hline

\end{tabular}
\end{center}
\end{table}

\begin{table}[t]
\begin{center}
\caption{Timing Performance Comparison}
\begin{tabular}{|l||l|l|l|l|}
\hline
{Problem $\backslash$ Method} & {1st-Euler} & {2nd-Euler} & {1st-RK4}& {2nd-RK4} \\ \hline
\textit{Cartpole}       &     0.020s       & 0.026s& 0.083s & 0.079s\\ \hline
\textit{Block  }      &    0.010s        & {0.010}s & 0.013s& 0.014s \\ \hline
\textit{Acrobot}       &     0.767s        & 0.519s & 3.301s & 3.427s\\ \hline
\textit{1D Quadrotor}       &     0.031s       & 0.032s & 0.051s & 0.057s\\ \hline
\textit{2D Quadrotor}       &     0.034s       & 0.038s & 0.052s & 0.063s\\ \hline
\textit{Five Links Biped} & 0.469s   &     0.500s       & 2.207s  & 2.123s\\ \hline

\end{tabular}
\end{center}
\end{table}
\subsubsection{Timing} To compare the run time performance, we measure the IPOPT solver time for each method. The initial guesses for the IPOPT solver were set to zeros for all problems and all methods to eliminate the effect from the initial guess. The experiments are conducted on a desktop computer equipped with an i7, 8-core 12th generation CPU at 2.10 GHz without GPU acceleration.

\subsection{Results}
TABLE I and TABLE II show accuracy results and timing performance results, respectively. The 2nd-Euler is more accurate than the 1st-Euler while having a similar computation time. It is also true for the 2nd-RK4 and the 1st-RK4. This indicates that our proposed methods increase the approximation accuracy by considering the inherent relationship between the system configuration and its time derivatives. Besides, among all the methods, the 2nd-RK4 provides the most accurate results because it is a higher-order transcription method and uses the proposed method to handle the second-order system dynamics constraints.

\begin{figure}[t]
	\centering
	\includegraphics[width=9cm]{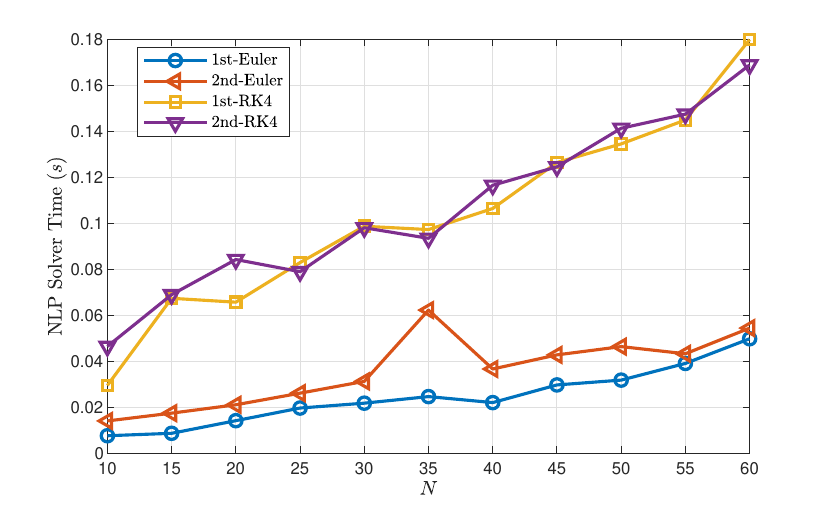}
	\caption{\small{The timing performance comparison of different methods in the cartpole swing-up problem in relation to the number of time intervals $N$.}}
	\label{fig:figure2}
\end{figure}
\begin{figure}[t]
	\centering
	\includegraphics[width=9cm]{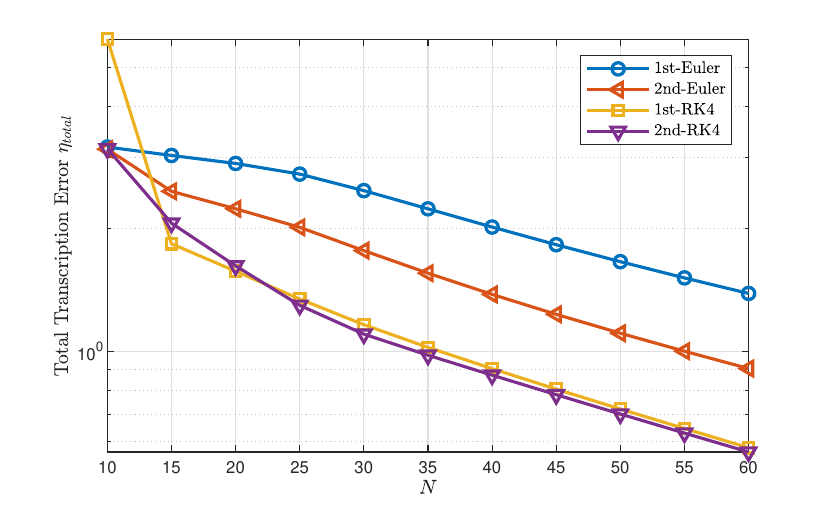}
	\caption{\small{The total transcription error comparison of different methods in the cartpole swing-up problem in relation to the number of time intervals $N$.}}
	\label{fig:figure3}
\end{figure}

We also evaluate the accuracy of the above four methods in relation to the number of time intervals $N$, as well as the timing performance in relation to the number of time intervals. The typical results of the cartpole swing-up problem are shown in Fig. \ref{fig:figure2} and Fig. \ref{fig:figure3}. Fig. \ref{fig:figure2} shows the result of the timing performance comparison, while Fig. \ref{fig:figure3} shows the result of the total transcription error versus the number of time intervals. The timing performance of the original methods and the modified methods are similar. It is clear that the modified Euler method has a significant improvement over the original Euler method, and the modified RK4 method performs best among the four methods. Furthermore, both the Euler method and the RK4 method converge as the number of time intervals increases. This confirms the theoretical results about convergence presented in \textit{Theorem 1} and \textit{Theorem 2}.

\section{Conclusion}
{In this paper, we studied numerical optimal control for high-order systems with the direct shooting method. We demonstrated the contradictory dynamics issue of the conventional direct shooting method when handling high-order systems and derived the detailed modified Euler and Runge-Kutta-4 methods for second-order systems. We also illustrated how to extend the proposed idea to high-order systems. Additionally, we proved the convergence properties of the proposed methods. Our methods were evaluated with several optimal control problems, which illustrated the superior performance of our methods. We are now working on extending the proposed methods to DDP-based algorithms, which can further enhance the advantage of direct shooting in numerical optimal control.}

\bibliographystyle{IEEEtran}
\bibliography{reference}

\begin{thebibliography}{10}
\providecommand{\url}[1]{#1}
\csname url@samestyle\endcsname
\providecommand{\newblock}{\relax}
\providecommand{\bibinfo}[2]{#2}
\providecommand{\BIBentrySTDinterwordspacing}{\spaceskip=0pt\relax}
\providecommand{\BIBentryALTinterwordstretchfactor}{4}
\providecommand{\BIBentryALTinterwordspacing}{\spaceskip=\fontdimen2\font plus
\BIBentryALTinterwordstretchfactor\fontdimen3\font minus
  \fontdimen4\font\relax}
\providecommand{\BIBforeignlanguage}[2]{{%
\expandafter\ifx\csname l@#1\endcsname\relax
\typeout{** WARNING: IEEEtran.bst: No hyphenation pattern has been}%
\typeout{** loaded for the language `#1'. Using the pattern for}%
\typeout{** the default language instead.}%
\else
\language=\csname l@#1\endcsname
\fi
#2}}
\providecommand{\BIBdecl}{\relax}
\BIBdecl

\bibitem{yang2022hierarchical}
B.~Yang, Y.~Lu, X.~Yang, and Y.~Mo, ``A hierarchical control framework for
  drift maneuvering of autonomous vehicles,'' in \emph{IEEE International
  Conference on Robotics and Automation}, 2022, pp. 1387--1393.

\bibitem{mpcc-uzh}
A.~Romero, S.~Sun, P.~Foehn, and D.~Scaramuzza, ``Model predictive contouring
  control for time-optimal quadrotor flight,'' \emph{IEEE Transactions on
  Robotics}, vol.~38, no.~6, pp. 3340--3356, 2022.

\bibitem{10124823}
R.~Wang, H.~Li, B.~Liang, Y.~Shi, and D.~Xu, ``Policy learning for nonlinear
  model predictive control with application to {USVs},'' \emph{IEEE
  Transactions on Industrial Electronics}, pp. 1--9, 2023.

\bibitem{betts2010practical}
J.~T. Betts, \emph{{Practical Methods for Optimal Control and Estimation Using
  Nonlinear Programming}}.\hskip 1em plus 0.5em minus 0.4em\relax SIAM, 2010.

\bibitem{kelly2017introduction}
M.~Kelly, ``An introduction to trajectory optimization: How to do your own
  direct collocation,'' \emph{SIAM Review}, vol.~59, no.~4, pp. 849--904, 2017.

\bibitem{Kelly_OptimTraj_Trajectory_Optimization_2022}
\BIBentryALTinterwordspacing
M.~P. Kelly, ``{OptimTraj: Trajectory Optimization for Matlab},'' 2022.
  [Online]. Available: \url{https://github.com/MatthewPeterKelly/OptimTraj}
\BIBentrySTDinterwordspacing

\bibitem{GPOPS-II}
M.~A. Patterson and A.~V. Rao, ``{GPOPS-II}: A {MATLAB} software for solving
  multiple-phase optimal control problems using {Hp}-adaptive {Gaussian}
  quadrature collocation methods and sparse nonlinear programming,'' vol.~41,
  no.~1, 2014.

\bibitem{becerra2010psopt}
V.~M. Becerra, ``{PSOPT} optimal control solver user manual,'' \emph{University
  of Reading}, 2010.

\bibitem{mayne1973differential}
D.~Q. Mayne, ``Differential dynamic programming--a unified approach to the
  optimization of dynamic systems,'' in \emph{Control and Dynamic
  Systems}.\hskip 1em plus 0.5em minus 0.4em\relax Elsevier, 1973, vol.~10, pp.
  179--254.

\bibitem{li2004iterative}
W.~Li and E.~Todorov, ``Iterative linear quadratic regulator design for
  nonlinear biological movement systems,'' in \emph{Proceedings of the First
  International Conference on Informatics in Control, Automation and Robotics},
  vol.~2.\hskip 1em plus 0.5em minus 0.4em\relax SciTePress, 2004, pp.
  222--229.

\bibitem{altro2019}
T.~A. Howell, B.~E. Jackson, and Z.~Manchester, ``{ALTRO}: A fast solver for
  constrained trajectory optimization,'' in \emph{IEEE/RSJ International
  Conference on Intelligent Robots and Systems (IROS)}, 2019, pp. 7674--7679.

\bibitem{OCS2}
F.~Farshidian \emph{et~al.}, ``{OCS2}: An open source library for optimal
  control of switched systems,'' [Online]. Available:
  \url{https://github.com/leggedrobotics/ocs2}.

\bibitem{PWA}
B.~E. Jackson, K.~Tracy, and Z.~Manchester, ``Planning with attitude,''
  \emph{IEEE Robotics and Automation Letters}, vol.~6, no.~3, pp. 5658--5664,
  2021.

\bibitem{junma_admm}
J.~Ma, Z.~Cheng, X.~Zhang, M.~Tomizuka, and T.~H. Lee, ``Alternating direction
  method of multipliers for constrained iterative {LQR} in autonomous
  driving,'' \emph{IEEE Transactions on Intelligent Transportation Systems},
  vol.~23, no.~12, pp. 23\,031--23\,042, 2022.

\bibitem{Alcan2023TrajectoryOO}
G.~Alcan, F.~J. Abu-Dakka, and V.~Kyrki, ``Trajectory optimization on matrix
  lie groups with differential dynamic programming and nonlinear constraints,''
  \emph{ArXiv}, vol. abs/2301.02018, 2023.

\bibitem{9811647}
W.~Jallet, N.~Mansard, and J.~Carpentier, ``Implicit differential dynamic
  programming,'' in \emph{International Conference on Robotics and Automation
  (ICRA)}, 2022, pp. 1455--1461.

\bibitem{moreno2022collocation}
S.~Moreno~Mart{\'\i}n, L.~Ros~Giralt, and E.~Celaya~Llover, ``Collocation
  methods for second order systems,'' in \emph{Proceedings of the XVIII
  Robotics: Science and Systems Conference (RSS)}, 2022, pp. 1--11.

\bibitem{simpson2022direct}
L.~Simpson, A.~Nurkanović, and M.~Diehl, ``{Direct collocation for numerical
  optimal control of second-order ODE},'' in \emph{Proceedings of European
  Control Conference (ECC)}, 2023, pp. 1--7.

\bibitem{moreno2022legendre}
S.~Moreno-Mart{\'\i}n, L.~Ros, and E.~Celaya, ``A {Legendre-Gauss}
  pseudospectral collocation method for trajectory optimization in second order
  systems,'' in \emph{Proceedings of IEEE/RSJ International Conference on
  Intelligent Robots and Systems (IROS)}, 2022, pp. 13\,335--13\,340.

\bibitem{burden2015numerical}
R.~L. Burden, J.~D. Faires, and A.~M. Burden, \emph{Numerical Analysis}.\hskip
  1em plus 0.5em minus 0.4em\relax Cengage learning, 2015.

\bibitem{andersson2019casadi}
J.~A. Andersson, J.~Gillis, G.~Horn, J.~B. Rawlings, and M.~Diehl, ``{CasADi}:
  a software framework for nonlinear optimization and optimal control,''
  \emph{Mathematical Programming Computation}, vol.~11, pp. 1--36, 2019.

\bibitem{wachter2006implementation}
A.~W{\"a}chter and L.~T. Biegler, ``On the implementation of an interior-point
  filter line-search algorithm for large-scale nonlinear programming,''
  \emph{Mathematical Programming}, vol. 106, pp. 25--57, 2006.

\bibitem{1234651}
C.~Chevallereau, G.~Abba, Y.~Aoustin, F.~Plestan, E.~Westervelt,
  C.~Canudas-De-Wit, and J.~Grizzle, ``{RABBIT: a testbed for advanced control
  theory},'' \emph{IEEE Control Systems Magazine}, vol.~23, no.~5, pp. 57--79,
  2003.

\end{thebibliography}
\end{document}